\documentclass[twocolumn,showpacs,superscriptaddress]{revtex4}%
\usepackage{amsfonts}
\usepackage{amsmath}
\usepackage{amssymb}
\usepackage{graphicx}%
\providecommand{\U}[1]{\protect\rule{.1in}{.1in}}

\begin{document}
\title{First-principles calculations of phase transition, elasticity, and thermodynamic properties for TiZr alloy}
\author{Bao-Tian Wang}
\affiliation{Institute of Theoretical Physics and Department of
Physics, Shanxi University, Taiyuan 030006, People's Republic of
China} \affiliation{LCP, Institute of Applied Physics and
Computational Mathematics, Beijing 100088, People's Republic of
China}
\author{Wei-Dong Li}
\affiliation{Institute of Theoretical Physics and Department of
Physics, Shanxi University, Taiyuan 030006, People's Republic of
China}
\author{Ping Zhang}
\thanks{Corresponding author; zhang\_ping@iapcm.ac.cn}
\affiliation{LCP, Institute of Applied Physics and Computational
Mathematics, Beijing 100088, People's Republic of China}
\affiliation{Center for Applied Physics and Technology, Peking
University, Beijing 100871, People's Republic of China}


\begin{abstract}
Structural transformation, pressure dependent elasticity behaviors,
phonon, and thermodynamic properties of the equiatomic TiZr alloy
are investigated by using first-principles density-functional
theory. Our calculated lattice parameters and equation of state for
$\alpha$ and $\omega$ phases as well as the phase transition
sequence of
$\alpha$$\mathtt{\rightarrow}$$\omega$$\mathtt{\rightarrow}$$\beta$
are consistent well with experiments. Elastic constants of $\alpha$
and $\omega$ phases indicate that they are mechanically stable. For
cubic $\beta$ phase, however, it is mechanically unstable at zero
pressure and the critical pressure for its mechanical stability is
predicted to equal to 2.19 GPa. We find that the moduli, elastic
sound velocities, and Debye temperature all increase with pressure
for three phases of TiZr alloy. The relatively large $B/G$ values
illustrate that the TiZr alloy is rather ductile and its ductility
is more predominant than that of element Zr, especially in $\beta$
phase. Elastic wave velocities and Debye temperature have abrupt
increase behaviors upon the $\alpha$$\mathtt{\rightarrow}$$\omega$
transition at around 10 GPa and exhibit abrupt decrease feature upon
the $\omega$$\mathtt{\rightarrow}$$\beta$ transition at higher
pressure. Through Mulliken population analysis, we illustrate that
the increase of the \emph{d}-band occupancy will stabilize the cubic
$\beta$ phase. Phonon dispersions for three phases of TiZr alloy are
firstly presented and the $\beta$ phase phonons clearly indicate its
dynamically unstable nature under ambient condition. Thermodynamics
of Gibbs free energy, entropy, and heat capacity are obtained by
quasiharmonic approximation and Debye model.
\end{abstract}
\maketitle

\section{INTRODUCTION}
Group IV transition metals such as titanium, zirconium, and hafnium
as well as their alloys have been investigated extensively by
experiments and theoretical calculations since their particular
applications in the aerospace, nuclear, and chemical industries
\cite{Ikehata,LiuPRB,Perez,Vohra}. Their properties of low thermal
neutron absorption cross section, adequate strength and ductility,
good corrosion resistance, long-term dimensional stability in an
irradiation environment, excellent compatibility with the fuel and
coolant, good thermal conductivity and adequate resistance to
fracture project them to the front of material applications for both
fuel element cans and in-core structural components in water-cooled
nuclear reactors \cite{Kutty}. Specifically, TiZr alloy is
traditionally used for high-pressure neutron diffraction and has
similar machining properties to the ZrNb alloy used in
nuclear-reactor tubing. Therefore, investigations of their
structural and elastic properties, which relate to various
fundamental solid-state properties such as interatomic potentials,
equation of state, phonon spectra, and thermodynamic properties,
have prominent meaning in both scientific and technological fields.
The scientific interest in Group IV transition metals and their
alloys origins from the fact that they have a narrow \emph{d}-band
in the midst of a broad \emph{sp}-band. The \emph{d}-band occupancy
is crucial for the structural stability of these materials and a
pressure-induced \emph{s-d} electron transfer is the driving force
behind their structural and electronic transitions
\cite{Duthie,Skriver}.

At ambient condition, both Ti and Zr crystallize in hexagonal
closed-packed (hcp) structure ($\alpha$ phase). At high temperature
of 1155 and 1135 K, they transform into the body-centered cubic
(bcc) $\beta$ phase, respectively. At room temperature and under
compression, $\alpha$-Zr transforms to another hexagonal structure
$\omega$ phase at about 2$-$7 GPa
\cite{Zilbershtein,Sikka,Xia1,ZhaoPRB}. Under further high pressure
of 30$-$35 GPa, the $\omega$$\mathtt{\rightarrow}$$\beta$ phase
transition has been observed
\cite{Xia1,Xia2,Akahama1,Akahama2,ZhaoAPL}. The whole
$\alpha$$\mathtt{\rightarrow}$$\omega$$\mathtt{\rightarrow}$$\beta$
transition series for Zr have been reproduced in theoretical
investigations \cite{WangZr,HaoPRB,Schnell,Ahuja}. However,
 the experimental established room
temperature transition sequence of Ti is
$\alpha$$\mathtt{\rightarrow}$$\omega$$\mathtt{\rightarrow}$$\gamma$$\mathtt{\rightarrow}$$\delta$
\cite{Vohra,Xia1,Akahama3,Errandonea}. The $\beta$ phase of Ti metal
has not been observed up to 216 GPa \cite{Akahama3}. Recent
first-principles study performed by Mei \emph{et al.} \cite{MeiPT}
found that the $\delta$-Ti is not stable under hydrostatic
compression and the 0 K phase transition sequence of Ti is
$\alpha$$\mathtt{\rightarrow}$$\omega$$\mathtt{\rightarrow}$$\gamma$$\mathtt{\rightarrow}$$\beta$.
The absence of the high-pressure $\beta$ phase for Ti in experiments
was attributed to the possible nonhydrostatic stress which distorts
the $\beta$ phase \cite{Verma}.

For Ti$-$Zr alloys, transition sequence of
$\alpha$$\mathtt{\rightarrow}$$\omega$$\mathtt{\rightarrow}$$\beta$
upon compression has been reported by Bashkin \emph{et al.}
\cite{BashkinPSS2000,BashkinJETP,BashkinPRB}. Using the differential
thermal analysis (DTA) and calorimetric technique, they extensively
studied the effects of temperature and pressure on the
$\alpha$$\mathtt{\rightleftarrows}$$\omega$ and
$\omega$$\mathtt{\rightleftarrows}$$\beta$ transformations in the
equiatomic TiZr alloy at temperatures up to 1023 K, and pressures up
to 7 GPa. \cite{BashkinPSS2000} At atmospheric pressure, the
$\beta$$\mathtt{\rightleftarrows}$$\alpha$ transition temperature
was measured to be 852 K upon cooling. Under pressure, the
transition temperature decreases down to the triple
$\alpha$$-$$\beta$$-$$\omega$ equilibrium point
($P_{\rm{tr}}$=4.9$\pm$0.3 GPa, $T_{\rm{tr}}$=733$\pm$30 K). At
pressures above the triple point, the slope of the
$\omega$$-$$\beta$ equilibrium boundary is positive at pressures up
to 7 GPa. At room temperature, the
$\alpha$$\mathtt{\rightleftarrows}$$\omega$ transition occurs at
around 5 GPa. Cooling of the $\beta$ phase in the pressure range
2.8$-$4.8 GPa can form a two-phase mixture of a stable $\alpha$ and
a metastable $\omega$ phase. Another group \cite{Aksenenkov}
reported that the equilibrium $\alpha$$-$$\omega$ boundary is
situated on the $P-T$ diagram at 6.6 GPa. Afterward, Bashkin
\emph{et al.} \cite{BashkinJETP,BashkinPRB} employed the
energy-dispersive X-ray diffraction (EDXD) with synchrotron
radiation to investigate the structural transitions of the TiZr
alloy. They found an increase in the
$\omega$$\mathtt{\rightarrow}$$\beta$ transition pressure from about
30 to 43$-$57 GPa when the titanium content in the alloys increases
from 0 to 50 at. \%. The pressure dependant behaviors of
superconductivity was also presented. For equiatomic TiZr alloy,
they found that the $\alpha$ phase remains the sole stable phase
under quasi-hydrostatic pressure up to 12.2 GPa \cite{BashkinPRB}.
Only from 15.5 GPa on, the $\omega$ phase becomes dominant. Recent
\emph{in situ} high-temperature high-pressure angle-dispersive
synchrotron radiation diffraction experiment revealed that under
pressure the equiatomic TiZr alloy occurs an
$\omega$$\mathtt{\rightarrow}$$\omega'$ isostructural transition at
high-pressure domain \cite{Dmitriev}.

However, to date only one theoretical study has focused on TiZr
alloy \cite{Trubitsin}, where they calculated the electron structure
and total energy by the scalar relativistic full-potential
linearized augmented-planewave (FPLAPW) method with a generalized
gradient approximation (GGA) potential. The $P-T$ phase diagram and
tendency toward decomposition in equiatomic TiZr alloy were
calculated within the electron density functional theory (DFT) and
the Debye-Gr$\ddot{u}$neisen model. Their calculations showed that
the $\omega$ phase is stable at atmospheric pressure and moderate
(up to 610 K) temperatures, and hence no
$\alpha$$\mathtt{\rightarrow}$$\omega$ transition occurs at room
temperature. They attributed the discrepancy between experiment and
theory to the experimental samples, within which defects may exist.
In present work, we have carefully calculated the structural, phase
transition, elastical, phonon vibrational, and thermodynamic
properties for equiatomic TiZr alloy by employing the
first-principles total energy calculations. Elastic constants,
elastic properties, and phonon dispersions of three experimentally
observed phases are firstly presented. Our main motivation is to
give out detail pressure behaviors of the elasticity as well as the
thermodynamic properties to support the practical application of
TiZr alloy in nuclear technology.

\section{computational methods}
First-principles DFT calculations on the basis of the frozen-core
projected augmented wave (PAW) method of Bl\"{o}chl \cite{PAW} are
performed within the Vienna \textit{ab initio} simulation package
(VASP) \cite{Kresse3}, where the Perdew, Burke, and Ernzerhof (PBE)
\cite{PBE} form of the GGA is employed to describe electron exchange
and correlation. To obtain accurate total energy and stress tensor,
a cutoff energy of 500 eV is used for the plane-wave set. The
$\Gamma$-centered \emph{k} point-meshes in the full wedge of the
Brillouin zone (BZ) are sampled by 18$\times $18$\times$16,
16$\times$16$\times$9, and 18$\times$18$\times$18 grids according to
the Monkhorst-Pack (MP) \cite{Monk} for $\alpha$ (two atoms cell),
$\omega$ (six atoms 1$\times$1$\times$2 supercell), and $\beta$ (two
atoms cell) phases TiZr alloy, respectively. Full geometry
optimization at each volume is considered to be completed when all
atoms are fully relaxed until the Hellmann-Feynman forces becoming
less than 0.001 eV/\AA. The Ti
3$s$$^{2}$3$p$$^{6}$3$d$$^{3}$4$s$$^{1}$ and the Zr
4$s$$^{2}$4$p$$^{6}$4$d$$^{3}$5$s$$^{1}$ orbitals are explicitly
included as valence electrons.

\section{results}
\subsection{Structure}
In calculations of $\alpha$ phase TiZr alloy, we use one lamella
structure, which has the same unit cell as the conventional hcp unit
cell but the Ti and Zr atoms are located on the layer along the
\emph{c} axis alternately. For $\beta$ phase, Ti atoms are fixed at
corner and Zr atoms are located at center of the bcc unit cells. For
$\omega$ TiZr, two types (A-type and B-type) of 1$\times$1$\times$2
supercells are constructed (Fig. 1). The theoretical equilibrium
structural parameters, bulk modulus \emph{B}, and pressure
derivative of the bulk modulus \emph{B$^{\prime}$} obtained by
fitting the energy-volume data in the third-order Birch-Murnaghan
equation of state (EOS) \cite{Birch} are tabulated in Table 1.
Experimental values from Ref. \cite{BashkinPRB} and other
theoretical results from Refs. \cite{Ikehata} and \cite{Trubitsin}
are also presented for comparison. Note that we only list in Table 1
the results of B-type structure for $\omega$ phase. Our calculated
equilibrium structural parameters for A-type $\omega$ phase are
$a_{0}$=4.829 {\AA} and $c_{0}$=3.013 {\AA}. Its bulk modulus
\emph{B} and pressure derivative of the bulk modulus
\emph{B$^{\prime}$} are equal to 100.6 GPa and 3.63, respectively.
All these values are almost equal to that of B-type structure.
However, the total energy of B-type structure is smaller by 0.0769
eV per formula unit than that of A-type structure. Using A-type
structure for $\omega$ phase will lead to incorrect results in
calculations of phase transition and mechanical properties. Thus,
the results of A-type $\omega$ phase are not presented in the
following and unless otherwise stated results of $\omega$ phase
refer to that of B-type structure.

\begin{figure}[ptb]
\begin{center}
\includegraphics[width=0.8\linewidth]{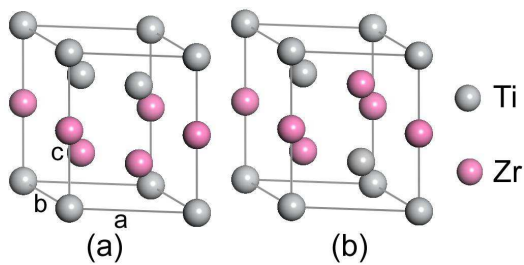}
\end{center}
\caption{(Color online) Schematic pictures of 1$\times$1$\times$2 supercells for $\omega$ phase in (a) A-type and (b) B-type structures.}%
\label{structure}%
\end{figure}

\begin{table*}[ptb]
\caption{Calculated lattice parameters (\emph{a} or \emph{c}), bulk
modulus (\emph{B}), pressure derivative of the bulk modulus
(\emph{B$^{^{\prime}}$}), and elastic constants of $\alpha$,
$\omega$, and $\beta$ TiZr at different pressures. For comparison,
experimental values and other theoretical results are also listed.}%
\label{elastic}
\begin{ruledtabular}
\begin{tabular}{cccccccccccccccccc}
Phase&Method&Pressure&a&c&\emph{B}&\emph{B$^{^{\prime}}$}&\emph{C$_{11}$}&\emph{C$_{12}$}&\emph{C$_{13}$}&\emph{C$_{33}$}&\emph{C$_{44}$}\\
&&(GPa)&({\AA})&({\AA})&(GPa)&&(GPa)&(GPa)&(GPa)&(GPa)&(GPa)\\
\hline
$\alpha$&This work&0&3.111&4.913&97.8&3.47&145.1&72.5&70.9&169.0&30.0\\
&&5&3.059&4.847&&&162.6&85.2&80.8&193.9&31.5\\
&&10&3.013&4.795&&&181.2&95.7&90.1&217.8&33.4\\
&DFT-PBE$^{\emph{a}}$&0&3.114&4.918&&&137.7&75.3&67.8&164.0&30.0\\
&Expt.$^{\emph{b}}$&0&3.104&4.923&148&3.8&&&&&\\
$\omega$&This work&0&4.825&3.011&100.5&3.55&170.9&76.2&51.9&209.2&39.6\\
&&10&4.697&2.923&&&216.1&97.6&66.7&257.2&45.6\\
&&20&4.592&2.860&&&257.7&120.1&80.6&301.4&49.5\\
&&30&4.507&2.808&&&298.5&141.7&93.5&340.9&51.4\\
&&40&4.438&2.762&&&336.8&163.7&105.3&375.5&51.6\\
&Expt.$^{\emph{b}}$&0&4.840&2.991&146&1.7&&&&&\\
$\beta$&This work&0&3.423&&94.8&3.35&91.5&96.8&&&36.6\\
&&10&3.321&&&&134.4&113.0&&&30.2\\
&&20&3.244&&&&172.9&129.2&&&38.9\\
&&30&3.181&&&&211.1&142.9&&&47.0\\
&&40&3.128&&&&249.9&153.1&&&55.0\\
&&50&3.081&&&&286.9&164.6&&&64.5\\
&&60&3.040&&&&322.6&176.3&&&76.0\\
&FPLAPW-GGA$^{\emph{c}}$&0&3.417&&&&&&&&\\
&Expt.$^{\emph{b}}$&57&3.098&&&&&&&&\\
\end{tabular}
$^{\emph{a}}$ Reference \cite{Ikehata}, $^{\emph{b}}$ Reference
\cite{BashkinPRB}, $^{\emph{c}}$ Reference \cite{Trubitsin}.
\end{ruledtabular}
\end{table*}

From Table 1, one can find excellent coincidence between our
calculated values and the corresponding experimental results for
$\alpha$ and $\omega$ structural parameters. But our calculated bulk
moduli \emph{B} for these two phases are largely smaller than that
from experiment. Although we can not simply attribute these results
to the under-binding effect of the GGA approach, we find that our
calculated bulk modulus \emph{B} for $\alpha$ TiZr alloy lies within
the range of experimental values between Ti 102 GPa \cite{Vohra} and
Zr 92 GPa \cite{ZhaoPRB}. As for $\beta$ phase, our calculated
equilibrium crystal constant is in good agreement with recent
theoretical calculation \cite{Trubitsin}, where a value of 3.417
\AA\ was given out. The discrepancy between our results and the
experimental value (3.417 \AA\ at 57 GPa) \cite{BashkinPRB} mainly
due to the temperature effect.

\begin{figure}[ptb]
\begin{center}
\includegraphics[width=0.8\linewidth]{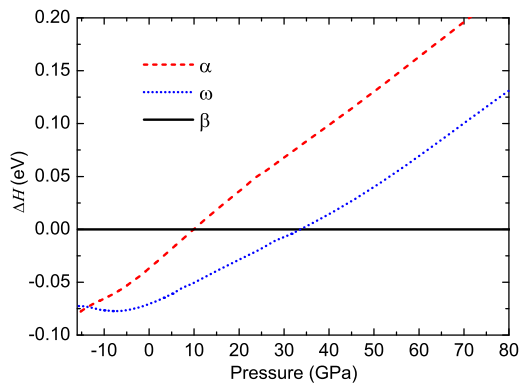}
\end{center}
\caption{(Color online) Calculated enthalpy differences of $\alpha$
and $\omega$ phases with respect to $\beta$ phase as a function of
pressure.}%
\label{enthalpy}%
\end{figure}

\begin{table}
\caption{Calculated transition pressure for TiZr. For comparison,
other theoretical works and available experimental data for TiZr,
Ti, and Zr are listed.} \label{mechanical}
\begin{tabular}{llcrccccccc}
\hline\hline&&\multicolumn{2}{c}{Transition pressure}\\
&&\multicolumn{2}{c}{(GPa)}\\
\cline{3-4}
Metal&Transition&Expt.&Theory\\
\hline
TiZr&$\alpha\rightarrow\omega$&5$-$11$^{\emph{a}}$&$-$13.8$^{\emph{b}}$\\
&$\omega\rightarrow\beta$&43$^{\emph{a}}$&26$^{\emph{c}}$, 33.9$^{\emph{b}}$\\
Ti&$\alpha\rightarrow\omega$&2$-$11.9$^{\emph{d}}$&52$^{\emph{e}}$, $-$3.7$^{\emph{f}}$\\
Zr&$\alpha\rightarrow\omega$&2$-$7$^{\emph{g}}$&4.8$^{\emph{h}}$, 3.3$^{\emph{i}}$, $-$3.7$^{\emph{j}}$\\
&$\omega\rightarrow\beta$&30$-$35$^{\emph{k}}$&32.4$^{\emph{j}}$\\\hline\hline
\end{tabular}\\[2pt]
$^{\emph{a}}$ Reference \cite{BashkinPSS2000,BashkinPRB,Aksenenkov},
$^{\emph{b}}$ Present work, $^{\emph{c}}$ References
\cite{Trubitsin}, $^{\emph{d}}$ References
\cite{Vohra,Xia1,Akahama3,Errandonea}, $^{\emph{e}}$ Reference
\cite{JonaPSS}, $^{\emph{f}}$ Reference \cite{MeiPT}, $^{\emph{g}}$
References \cite{Sikka,Xia1,ZhaoPRB}, $^{\emph{h}}$ Reference
\cite{JomardZr}, $^{\emph{i}}$ Reference \cite{LandaZr},
$^{\emph{j}}$ Reference \cite{WangZr}, $^{\emph{k}}$ References
\cite{Xia1,Xia2,Akahama1,Akahama2,ZhaoAPL}.
\end{table}

\subsection{Phase transition at 0 K}
At 0 K, the Gibbs free energy is equal to the enthalpy \emph{H}.
After calculation, we can plot in Fig. 2 the relative enthalpies of
the $\alpha$ and the $\omega$ phases with respect to the $\beta$
phase as a function of pressure. The crossing between the $\alpha$
and $\omega$ enthalpy curves readily gives phase transition pressure
of $-$13.8 GPa, which indicates that at ambient pressure the
$\omega$ phase is more stable than the $\alpha$ phase. This fact is
in disagreement with experiment
\cite{BashkinPSS2000,BashkinPRB,Aksenenkov}, but coincides with
recent theoretical calculations \cite{Trubitsin}. Trubitsin \emph{et
al.} \cite{Trubitsin} attributed the discrepancy between experiment
and theory to the experimental impure samples. They presented
detailed explanation for this discrepancy. In our opinion, the
imperfect crystals used in experiments may responsible mainly for
this discrepancy. After all, the metastable $\omega$ phase was
obtained at atmospheric pressure through cooling of the $\beta$
phase under a pressure of 6 GPa with subsequent unloading at room
temperature. One other reason may from the temperature contribution.
In Table 2, we list the transition pressures for TiZr alloy as well
as pure Ti and Zr metals from experiments and theoretical
calculations. For the stable phase of metals Ti and Zr at ambient
pressure, there exist debates in theoretical studies
\cite{WangZr,HaoPRB,MeiPT,JomardZr,LandaZr,Ostanin,JonaJPCM,JonaPSS,GradPRB,GreefZr}
although experiments \cite{Vohra,Xia1,Akahama3,Sikka,ZhaoPRB} have
reported the most stable phase to be $\alpha$ phase. For metal Zr,
two DFT works \cite{JomardZr,LandaZr} using full-potential linear
muffin-tin orbital (FPLMTO)-GGA have illustrated the
$\alpha$$\mathtt{\rightarrow}$$\omega$ transition for Zr at 4.8 and
3.3 GPa, respectively. However, two other DFT works
\cite{Ostanin,JonaJPCM} using FPLMTO-GGA and FPLAPW-LDA methods
indicated that the most stable phase at 0 K under ambient pressure
is $\omega$, not $\alpha$ phase. Two recent DFT-PBE works
\cite{WangZr,HaoPRB} also support this conclusion. For metal Ti, the
debates also exist in DFT calculations \cite{MeiPT,JonaPSS} as
indicated in Table II. By adding the phonon contribution to the
Gibbs energy, recent DFT-PBE \cite{HaoPRB,Mei} studies explicitly
showed that the transition of $\alpha$$\mathtt{\rightarrow}$$\omega$
for Ti (Zr) occurs at 1.8 (1.7) GPa at room temperature. Thus, the
disagreement between some theoretical works and experiments for
elemental Ti and Zr mainly originates from temperature effect. The
entropy from the thermal population of phonon states will stabilize
the $\alpha$ phase at room temperature \cite{MeiPT}. In our present
work, we should also consider the temperature effect on the
$\alpha$$\mathtt{\rightarrow}$$\omega$ transition pressure for TiZr
alloy since the same scheme used with respect to our previous study
of Zr \cite{WangZr}. In addition, Trubitsin \emph{et al.}
\cite{Trubitsin} observed that $\alpha$ TiZr alloy is stable in the
temperature range 600 K$<$T$<$900 K after including lattice energy
and entropy in the Debye model. Therefore, the disagreement of
$\alpha$$\mathtt{\rightarrow}$$\omega$ transition pressure between
our calculation and experiment for TiZr alloy can also be attributed
mainly to the experimental impure samples and partially to the
temperature effect. As for the $\omega$$\mathtt{\rightarrow}$$\beta$
transition, our calculated transition pressure (33.9 GPa) is in good
agreement with experiment \cite{BashkinPRB}.

\begin{figure}[ptb]
\begin{center}
\includegraphics[width=0.8\linewidth]{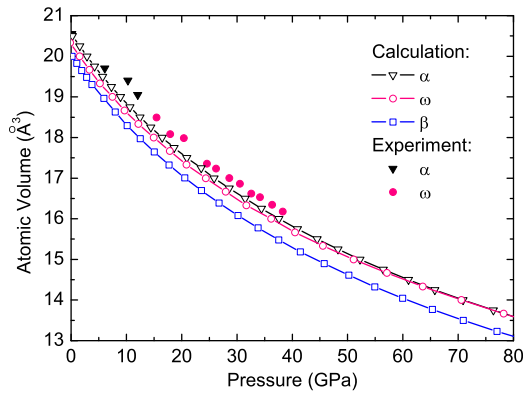}
\end{center}
\caption{(Color online) Calculated compression curves of TiZr alloy
compared with experimental measurements from Ref. [25].}%
\label{compression}%
\end{figure}

\begin{figure}[ptb]
\begin{center}
\includegraphics[width=0.8\linewidth]{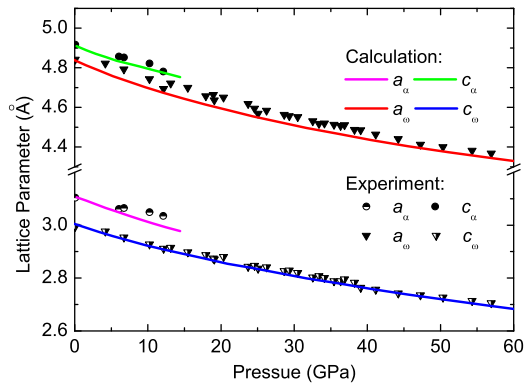}
\end{center}
\caption{(Color online) Calculated lattice parameters in comparison
with the experimental data of Bashkin \emph{et al.}
\cite{BashkinPRB} for $\alpha$ and $\omega$
phases of TiZr alloy. Our calculated results of $\omega{^{\prime}}$ phase are also presented.}%
\label{ac}%
\end{figure}

To further analyze the pressure behavior of TiZr alloy, we plot in
Fig. 3 the compression curves of $\alpha$, $\omega$, and $\beta$
phases. For comparison, the experimental data from Ref.
\cite{BashkinPRB} are also shown in the figure. Clearly, our
calculated \emph{P$-$V} equation of states are well consistent with
the experimental measurement. The slight smallness of our calculated
volumes compared to that of the experimental data can be attributed
to the temperature effect. The experiment was performed at room
temperature, while our data are valid only at 0 K. In Fig. 4, we
compare the calculated lattice parameters of $\alpha$ and $\omega$
phases with experimental results \cite{BashkinPRB}. Table 1 also
lists the optimized structural parameters of TiZr alloy at some
different pressures. In the whole pressure domain of 0$-$60 GPa, the
predicted lattice parameters for both $\alpha$ and $\omega$ phases
compare well with the experimental values, which supplies the
safeguard for our following study of mechanical and elastic
properties of TiZr alloy under pressure.

\subsection{Elasticity}
\begin{figure}[ptb]
\begin{center}
\includegraphics[width=0.8\linewidth]{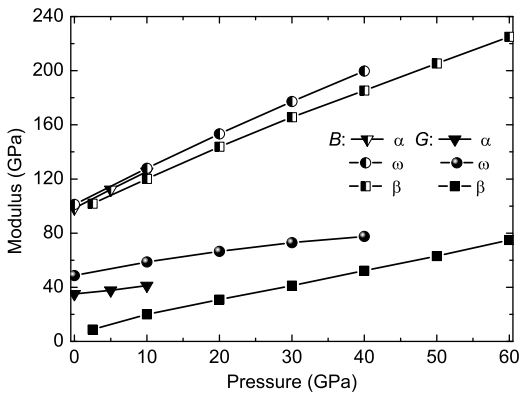}
\end{center}
\caption{Calculated moduli for TiZr alloy as a
function of pressure. Solid lines are guides to the eyes.}%
\label{BG}%
\end{figure}
Elastic constants can measure the resistance and mechanical features
of crystal to external stress or pressure, thus describing the
stability of crystals against elastic deformation. In present work,
elastic constants for TiZr alloy in $\alpha$, $\omega$, and $\beta$
phases at different pressures are calculated through applying stress
tensors with various small strains onto the optimized structures.
The strain amplitude $\delta$ is varied in steps of 0.006 from
$\delta$=$-$0.036 to 0.036. Results are presented in Table 1.
Obviously, the mechanical stability of $\alpha$ and $\omega$ TiZr at
0 GPa and at some typical finite pressures can be predicted from the
elastic constants data. But the elastic constants of $\beta$ phase
at 0 GPa illustrate that the cubic phase is mechanically unstable.
We notice that the $\beta$ phase metals Ti and Zr are also
mechanically unstable at 0 GPa \cite{Hu,WangZr}. Along with the
increase of pressure from 0 GPa to 60 GPa, the value of
$C_{11}-C_{12}$ increases near linearly from $-$5.3 GPa to 146.3 GPa
for TiZr. Fitting the curve (not shown) of the pressure behavior of
$C_{11}-C_{12}$ by polynomial function, we find that the value of
$C_{11}-C_{12}$ becomes positive at $P\geq$2.19 GPa. In fact, at
$P$=2.5 GPa our calculated $C_{11}, C_{12}$, and $C_{44}$ equal to
102.9, 101.2, and 25.6 GPa, respectively, which explicitly indicate
the elastically stable of bcc TiZr under this pressure.

\begin{figure}[ptb]
\begin{center}
\includegraphics[width=0.8\linewidth]{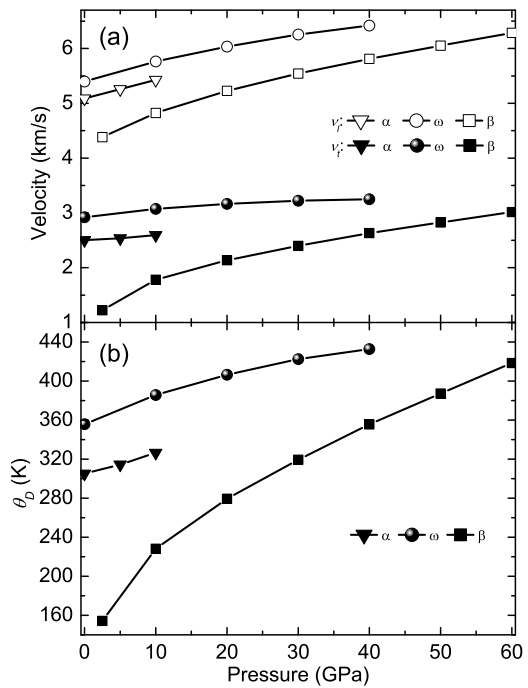}
\end{center}
\caption{Transverse and longitudinal sound velocities (a) and Debye
temperature (b) for TiZr alloy as a
function of pressure. Solid lines are guides to the eyes.}%
\label{DebyeT}%
\end{figure}

After obtaining elastic constants, the polycrystalline bulk modulus
\emph{B} and shear modulus \emph{G} are calculated from the
Voigt-Reuss-Hill (VRH) approximations \cite{Hill}. Results of TiZr
alloy in $\alpha$, $\omega$, and $\beta$ phases are shown in Fig. 5.
The deduced bulk moduli from VRH approximations for all three phases
at 0 GPa turn out to be very close to that obtained from the EOS
fitting, which indicates that our calculations are consistent and
reliable. Along with the increasing of pressure, both bulk modulus
\emph{B} and shear modulus \emph{G} increase near linearly for all
three phases. The increasing rates of \emph{B} for all three phases
are apparently larger than that of \emph{G}. It is well known that
the shear modulus $G$ represents the resistance to plastic
deformation, while the bulk modulus $B$ can represent the resistance
to fracture. A high (low) $B/G$ value is responsible for the
ductility (brittleness) of polycrystalline materials. The critical
value to separate ductile and brittle materials is about 1.75
\cite{Pugh}. Using the calculated values of bulk modulus $B$ and
shear modulus $G$ for TiZr alloy, the $B/G$ values for $\alpha$
phase increase from 2.80 to 3.04 under pressure from 0 GPa to 10
GPa, for $\omega$ phase increase from 2.08 to 2.57 upon compression
from 0 GPa to 40 GPa, and for $\beta$ phase decrease from 11.57 to
3.01 under pressure from 2.5 GPa to 60 GPa. These results indicate
that transit to $\omega$ phase will lead to less ductility and
further transit to $\beta$ phase will result in more predominance of
ductility. Although the large $B/G$ values for $\beta$ TiZr at
pressure range of 0$-$30 GPa can not be achieved in experiment at
room temperature, the $B/G$ value of about 3.5 for $\beta$ phase at
40 GPa indicates that the $\beta$ phase TiZr alloy possess the
biggest ductility. For $\alpha$ Zr, the $B/G$ value of 2.63 at
ambient condition can be derived from the elastic data in Ref.
\cite{LiuJAP}. Therefore, TiZr alloy is rather ductile and its
ductility is more predominant than that of element Zr. We hope that
our calculated elastic constants and elastic moduli can be
illustrative in the realistic application of the mechanical data for
TiZr alloy.

Elastic wave velocities provide important information about the
behavior of materials before phase transition. The transverse and
longitudinal elastic wave velocities of the polycrystalline
materials can be calculated through relations of
$\upsilon_{t}=\sqrt{G/\rho}$ ($\rho$ is the density) and
$\upsilon_{l}=\sqrt{(3B+4G)/3\rho}$, respectively. The average wave
velocity in the polycrystalline materials is approximately given as
$\upsilon_{m}=[(1/3)(2/\upsilon_{t}^{3}+1/\upsilon_{l}^{3})]^{-1/3}$.
Using the relation
$\theta_{D}=(h/k_{B})(3n/4\pi\Omega)^{1/3}\upsilon_{m}$, the Debye
temperature ($\theta_{D}$) can be obtained. The calculated results
of transverse and longitudinal elastic wave velocity are plotted in
Fig. 6(a) and the Debye temperature in Fig. 6(b). Increasing
behaviors of elastic wave velocities and $\theta_{D}$ under pressure
are obvious. Transiting from $\alpha$ to $\omega$ at around 10 GPa
(experimental transition pressure), TiZr alloy exhibit abrupt
increasing feature for both transverse and longitudinal elastic wave
velocities: from 5.43 to 5.76 km/s for $\upsilon_{t}$ and from 2.60
to 3.07 km/s for $\upsilon_{l}$. In study of element Zr, Liu
\emph{et al.} \cite{LiuJAP} observed similar behaviors of elastic
wave velocities by using ultrasonic interferometry in conjunction
with synchrotron x-ray radiation. From Fig. 6(a), we find that the
elastic wave velocities will decrease upon the $\omega$ to $\beta$
transition at high pressure. The Debye temperature also shows this
kind of increase first (at $\alpha$ to $\omega$ transition) and
decrease later (at $\omega$ to $\beta$ transition) behavior upon
compression.

\begin{figure}[ptb]
\begin{center}
\includegraphics[width=0.8\linewidth]{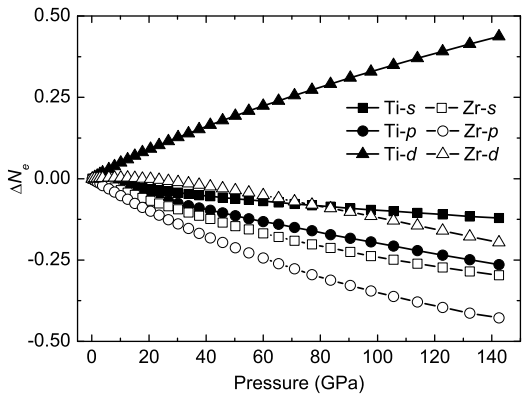}
\end{center}
\caption{Number of electrons on \emph{s}, \emph{p}, and \emph{d}
orbitals for TiZr alloy as a function of pressure, relative to their
values at 0 pressure. The numbers of \emph{s} electrons ($N_{s}$),
\emph{p} electrons ($N_{p}$), and \emph{d} electrons ($N_{d}$) at
\emph{P}=0 GPa for Ti atom
are 2.469, 6.417, and 3.012, respectively, for Zr atom are 2.565, 6.413, and 3.012, respectively.}%
\label{electron}%
\end{figure}

For discussion of pressure induced $s$-$d$ electron transfer for
TiZr alloy, we have performed Mulliken population analysis
\cite{Mulliken} of its $\beta$ phase and then plot in Fig. 7 the
variation of the number of electrons on \emph{s}, \emph{p}, and
\emph{d} orbitals with increasing pressure. Clearly, only the
$N_{d}$ of Ti atom increases with pressure. Arithmetic average of
$N_{d}$ for Ti and Zr atoms also possesses obvious
pressure-dependent increasing behavior. Increase of \emph{d}-band
occupancy will stabilize $\beta$ phase of TiZr alloy under pressure.
However, we find that the \emph{sp} electrons for Ti atom and the
\emph{spd} electrons for Zr atom all decrease upon compression up to
143 GPa. No increasing behavior of $N_{s}$ has been observed. This
fact is different from Ti \cite{Hu} and Zr \cite{Zhang}, where it
was found that the $N_{s}$ in both of these two elemental metals
start to increase when the pressure exceeds about 70 and 100 GPa,
respectively. Thus, the $s$-$d$ electron transfer behaviors for TiZr
alloy are different from its archetype metals.

\subsection{Phonon dispersion}
To our knowledge, although plenty of attentions have been paid on
the phonon dispersions, thermodynamics, and $P-T$ phase diagrams of
Ti \cite{Mei,Souvatzis} and Zr \cite{WangZr,HaoPRB,Schnell,Heiming},
no experimental and theoretical phonon dispersion results have been
published for equiatomic TiZr alloy. In this subsection and the
following subsection, we want to present our calculated results of
phonon dispersion curves, thermodynamics, and $P-T$ phase diagram
for TiZr alloy.

We use the supercell approach \cite{Parlinski} and the small
displacement method as implemented in the FROPHO code \cite{fropho}
to calculate the phonon curves in the BZ and the corresponding
phonon density of states (DOS). In the interpolation of the force
constants for the phonon dispersion curve calculations,
5$\times$5$\times$5, 3$\times $3$\times$1, and 5$\times$5$\times$5
\emph{k}-point meshes are used for $\alpha$ 3$\times$3$\times$3,
$\omega$ 3$\times$3$\times$3, and $\beta$ 4$\times$4$\times$4
supercells, respectively. The forces induced by small displacements
are calculated by VASP.

The calculated phonon dispersion curves and the phonon DOS of
$\alpha$, $\omega$, and $\beta$ TiZr alloy at 0 GPa are displayed in
Fig. 8. Overall speaking, behaviors of the phonons in the BZ for
$\alpha$ and $\omega$ TiZr alloy are similar to that of metal Ti
\cite{Mei} and Zr \cite{WangZr}. Same with Zr, the $\omega$ phonons
for TiZr alloy are stiffer along the $c$ axis than in the basal
plane due to the low $c/a$ ratio. The $\alpha$ phonons along [001]
direction for TiZr alloy are softer than that of metal Ti and Zr.
Compared with $\beta$ Ti and $\beta$ Zr, phonons in the BZ for
$\beta$ phase of TiZr alloy exhibit different behaviors. The
unstable phonon branch along [110] direction in the two metals
transfer to show stable and stiff feature in the alloy. On the
contrary, the phonons along [001] direction are stable for metals
but unstable for the alloy. These informations are important in
analyzing phase transformation. The stable nature of phonon branch
along [110] direction for $\beta$ TiZr alloy may responsible for the
difficulties of $\alpha$$\mathtt{\rightarrow}$$\beta$ transition
under high temperature and the $\omega$$\mathtt{\rightarrow}$$\beta$
transition under high pressure as indicated in experiments
\cite{BashkinPSS2000,BashkinPRB,Aksenenkov}.

\begin{figure}[ptb]
\begin{center}
\includegraphics[width=0.8\linewidth]{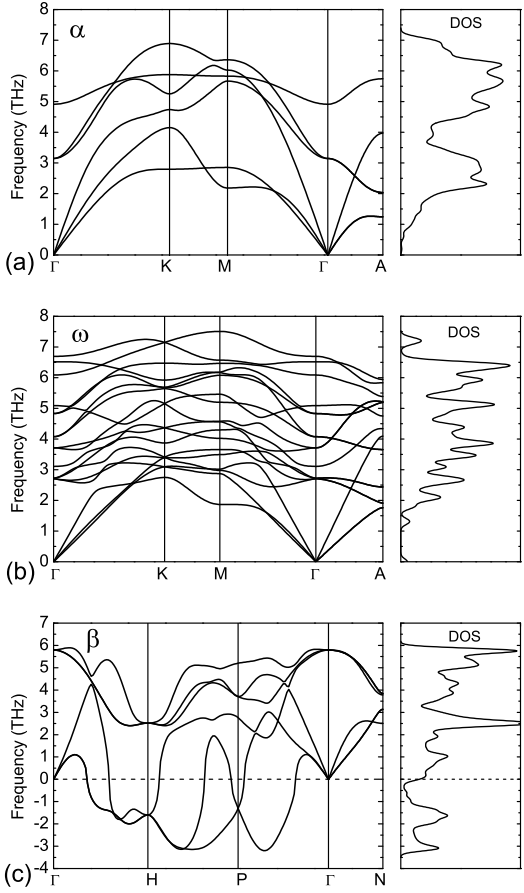}
\end{center}
\caption{Phonon dispersion curves and phonon DOS of $\alpha$, $\omega$, and $\beta$ TiZr alloy.}%
\label{phonon}%
\end{figure}

\subsection{Thermodynamic properties and $P-T$ phase diagram}
\begin{figure*}[ptb]
\begin{center}
\includegraphics[width=0.8\linewidth]{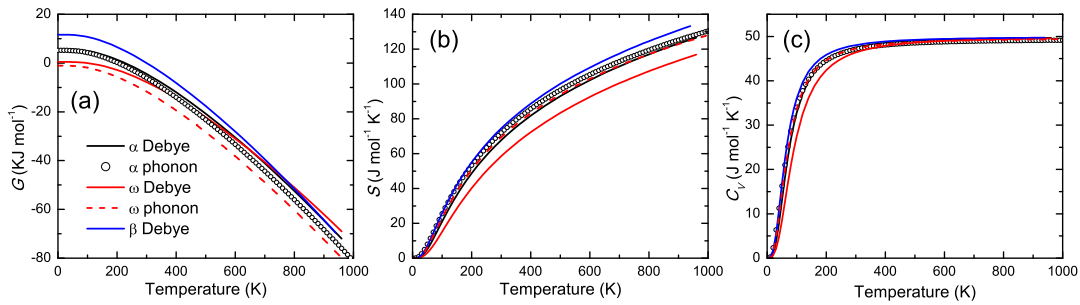}
\end{center}
\caption{Temperature dependences of (a) Gibbs free energy, (b)
entropy, and (c) specific heat at constant volume ($C_{V}$) for
$\alpha$, $\omega$,
and $\beta$ phases of TiZr alloy.}%
\label{thermal}%
\end{figure*}

Thermodynamic properties of equiatomic TiZr alloy can be determined
by phonon calculation using the quasiharmonic approximation (QHA)
\cite{Siegel,Zhang2010} or by the quasiharmonic Debye model
\cite{GIBBS}. Within these two models, the Gibbs free energy
\emph{G(T,P)} is written as
\begin{equation}
G(T,P)=F(T,V)+PV.
\end{equation}
Here, \emph{F(T,V)} is the Helmholtz free energy at temperature
\emph{T} and volume \emph{V} and can be expressed as
\begin{equation}
F(T,V)=E(V)+F_{vib}(T,V)+F_{el}(T,V),
\end{equation}
where \emph{E(V)} is the ground-state total energy,
\emph{F$_{vib}$(T,V)} is the vibrational energy of the lattice ions
and \emph{F$_{el}$(T,V)} is the thermal electronic contribution.

Under QHA, the \emph{F$_{vib}$(T,V)} can be calculated by
\begin{equation}
F_{vib}(T,V)=k_{B}T\int_{0}^{\infty}g(\omega)\ln\left[  2
\sinh\left( \frac{\hslash\omega}{2k_{B}T}\right)  \right]  d\omega,
\end{equation}
where $\omega$ represents the phonon frequencies and $g(\omega)$ is
the phonon DOS. This formula requests positive results of phonon
DOS. So it is not suitable for dynamically unstable phases. Instead,
the vibration energy for phases with imaginary phonon frequencies
can be estimated by the Debye model
\begin{equation}
F_{vib}(T,V)=\frac{9}{8}k_{B}\theta_{D}+k_{B}T\left[3\ln\left
(1-e^{-\frac{\theta_{D}}{T}}\right) -D\left
(\frac{\theta_{D}}{T}\right)\right],
\end{equation}
where $\frac{9}{8}k_{B}\theta_{D}$ is zero-point energy due to
lattice ion vibration at 0 K and $D(\theta_{D}/T)$ the Debye
integral written as
$D(\theta_{D}/T)=3/(\theta_{D}/T)^{3}\int_{0}^{\theta_{D}/T}x^{3}/(e^{x}-1)dx$.
Detailed computation scheme of Debye model please see Ref.
\cite{GIBBS}.

$F_{el}$ in Eq. (2) can be obtained from the energy and entropy
contributions, i.e., $E_{ele}-TS_{ele}$. The electronic entropy
$S_{ele}$ is of the form
\begin{equation}
S_{ele}(V,T)=-k_{B}\int n(\varepsilon,V)[f\ln
f+(1-f)\ln(1-f)]d\varepsilon,
\end{equation}
where $n(\varepsilon)$ is electronic DOS and $f$ is the Fermi-Dirac
distribution. The energy $E_{ele}$ due to the electron excitations
takes the following form:
\begin{equation}
E_{ele}(V,T)=\int n(\varepsilon,V)f\varepsilon
d\varepsilon-\int^{\varepsilon_{F}}n(\varepsilon,V)\varepsilon
d\varepsilon,
\end{equation}
where $\varepsilon_{F}$ is the Fermi energy.

Using QHA and Debye model, we calculated the Gibbs free energy
($G$), entropy ($S$), and specific heat at constant volume ($C_{V}$)
for $\alpha$, $\omega$, and $\beta$ phases of TiZr alloy at their
equilibrium volumes. As shown in Fig. 9, the Gibbs free energy and
entropy calculated by QHA are almost identical to those obtained by
employing the Debye model for $\alpha$ phase, while the differences
between these two schemes for $\omega$ phase is slightly large,
especially for high-temperature domain. Although the phonon
calculation is more reliable in general analysis, some nonharmonic
terms, which is important for $\omega$ phase, possibly having been
ignored in the QHA method. From Fig. 9(a), cross between $\alpha$
and $\omega$ curves by Debye calculations readily gives phase
transition temperature of 680 K. A phase transition temperature of
940 K for $\alpha$$\mathtt{\rightarrow}$$\beta$ also can be seen.
So, we can use Debye model to obtain the $P-T$ phase diagram by
changing the pressure $P$ (see Fig. 10). However, there is no cross
between $\alpha$ and $\omega$ curves using phonon calculations. This
need more works to clarify. Figure 9(b) shows the entropy results.
While the entropy of $\omega$ phase is lower than $\alpha$ phase,
the results of $\beta$ phase are higher than $\alpha$ phase in all
considered temperature domain. The specific heat at constant volume
($C_{V}$) can be directly calculated by $T(\frac{\partial
S}{\partial T})_{V}$. Results are presented in Fig. 9(c), using
which one can estimate the Debye temperature. Actually, our phonon
calculations indicate that the zero-point energies for $\alpha$ and
$\omega$ phases at $P$=0 GPa are 27.02 and 28.00 meV, respectively.
So, the Debye temperature values from phonon calculations equal to
278.6 and 288.9 K, respectively. The Debye model gives
$\theta_{D}$=303.6 and 369.0 K for $\alpha$ and $\omega$ phases,
respectively. So, the Debye model gives closer values than the
phonon calculations with respect to the Debye temperature values of
305.1 and 355.6 K computed by elastic constants (see Fig. 7).

\begin{figure}[ptb]
\begin{center}
\includegraphics[width=0.8\linewidth]{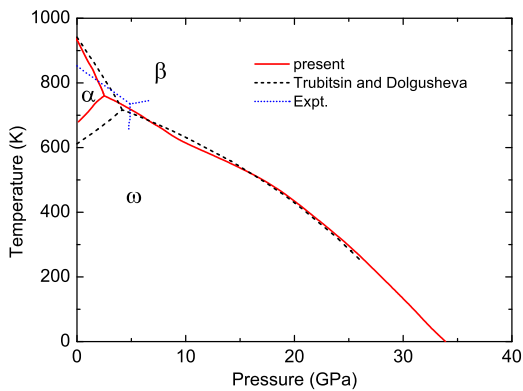}
\end{center}
\caption{Calculated $P-T$ phase diagram of TiZr alloy. For
comparison, we also plot in figure the experimental data
\cite{BashkinPSS2000} and the theoretical results from Ref.
\cite{Trubitsin}.}%
\label{phonon}%
\end{figure}

We present in Fig. 10 the $P-T$ phase diagram of equiatomic TiZr
alloy calculated by the Debye model. A triple point
($P_{\rm{tr}}$=2.5 GPa, $T_{\rm{tr}}$=760 K), comparable to
experimental results ($P_{\rm{tr}}$=4.9$\pm$0.3 GPa,
$T_{\rm{tr}}$=733$\pm$30 K) \cite{BashkinPSS2000} and previous
theoretical values ($P_{\rm{tr}}$=4.2 GPa, $T_{\rm{tr}}$=720 K)
\cite{Trubitsin}, is predicted. The
$\omega$$\mathtt{\rightarrow}$$\alpha$ and
$\alpha$$\mathtt{\rightarrow}$$\beta$ transition temperatures of 680
and 940 K are close to previous Debye-Gr\"{u}neisen values (610 and
943 K, respectively) \cite{Trubitsin}. Detailed understanding of the
$P-T$ phase diagram please see Ref. \cite{Trubitsin}.

\section{CONCLUSION}
In summary, structure, phase transition, elasticity, phonon, and
thermodynamics of the equiatomic TiZr alloy have been studied by
means of the first-principles DFT-PBE method. The calculated lattice
parameters and equations of state for $\alpha$ and $\omega$ phases
are in good agreement with experiments. We have found that the
$\omega$ phase is most stable at 0 GPa and the
$\omega$$\mathtt{\rightarrow}$$\beta$ transition was calculated to
occur at around 33.9 GPa. The elastic constants of $\alpha$ phase
TiZr alloy at $P$=0 GPa coincide well with previous calculations.
Mechanical stability of $\alpha$ and $\omega$ phases were predicted
and the $\beta$ phase is mechanically unstable at zero pressure. Our
elastic constants indicate that the $\beta$ phase will become
mechanically stable upon compression. Under pressure, both bulk
modulus \emph{B} and shear modulus \emph{G} increase near linearly
for all three phases. Increasing behaviors of elastic wave
velocities and Debye temperature under pressure have also been
predicted. It was found that elastic wave velocities and Debye
temperature increase first (at $\alpha$ to $\omega$ transition) and
decrease later (at $\omega$ to $\beta$ transition) abruptly upon
compression. The Mulliken population analysis showed that either
\emph{sp} electrons of Ti atom or \emph{spd} electrons of Zr atom
transfer to the Ti-\emph{d} orbital with increasing pressure.

In phonon dispersion study, the dynamic stable nature for $\alpha$
and $\omega$ phases are observed. The unstable modes of $\beta$
phase TiZr alloy along [001] and [111] directions have been shown
and these imaginary vibration branches are different from elemental
Ti and Zr. Based on phonon calculations and also the Debye model,
thermodynamic properties of Gibbs free energy, entropy, and specific
heat at constant volume as well as the $P-T$ phase diagram for TiZr
alloy have been presented.

\begin{acknowledgments}
We thank Yu Yang for useful discussion. This work was supported by
NSFC under Grant Nos. 11104170, 51071032, and 11074155, the
Foundations for Development of Science and Technology of China
Academy of Engineering Physics under Grant No. 2009B0301037.
\end{acknowledgments}

\end{document}